\documentclass[12pt,a4paper]{article}
\usepackage{a4wide}
\usepackage{latexsym}
\usepackage{epsf}
\usepackage{amssymb}
\begin{document}
\def\be{\begin{equation}}
\def\ee{\end{equation}}
\def\bea{\begin{eqnarray}}
\def\eea{\end{eqnarray}}

\def\pd{\partial}
\def\a{\alpha}
\def\b{\beta}
\def\g{\gamma}
\def\d{\delta}
\def\m{\mu}
\def\n{\nu}
\def\t{\tau} 
\def\l{\lambda}
\def\s{\sigma}
\def\e{\epsilon}
\def\scri{\mathcal{J}}
\def\cM{\mathcal{M}}
\def\tcM{\tilde{\mathcal{M}}}
\def\RR{\mathbb{R}}
\def\CC{\mathbb{C}}

\hyphenation{re-pa-ra-me-tri-za-tion}
\hyphenation{trans-for-ma-tions}


\begin{flushright}
IFT-UAM/CSIC-01-34\\
gr-qc/0111031\\
\end{flushright}

\vspace{1cm}

\begin{center}

{\bf\Large Are the String and Einstein Frames Equivalent?}

\vspace{.5cm}

{\bf Enrique \'Alvarez and Jorge Conde}
\footnote{E-mail: {\tt enrique.alvarez,jorge.conde@uam.es}}
\vspace{.3cm}

\vskip 0.4cm

{\it  Instituto de F\'{\i}sica Te\'orica, C-XVI,
  Universidad Aut\'onoma de Madrid
  E-28049-Madrid, Spain}\footnote{Unidad de Investigaci\'on Asociada
  al Centro de F\'{\i}sica Miguel Catal\'an (C.S.I.C.)}\\and\\
 {\it Departamento de F\'{\i}sica Te\'orica, C-XI,
  Universidad Aut\'onoma de Madrid \\
  E-28049-Madrid, Spain }

\vskip 0.2cm

\vskip 1cm


{\bf Abstract}

 The low energy physics as predicted by strings can be expressed in two
(conformally related) different variables, usually called {\em frames}.
The problem is raised as to whether it is physically possible in some 
situations to tell 
one from the other.

\end{center}


\newpage

\setcounter{page}{1}
\setcounter{footnote}{1}

\section{Introduction}
In the low energy (that is, wavelengths much longer than the string scale, $l_s
\equiv (\alpha^{\prime})^{1/2}$ ) limit all string theories predict at
least a common sector of massless fields (in any dimension): one associated 
with the 
{\em graviton}
$g_{\m\n}$, another with a scalar field $\phi$, called the {\em dilaton}\footnote{
Which we shall choose as dimensionless.}, which is 
related to the string coupling constant, $g_s$ in the sense that a 
{\em constant} variation of the dilaton, $\delta \phi$ produces a
corresponding variation in the coupling constant 
$\frac{\delta g_s}{g_s}=\delta\phi$, and a two-
index antisymmetric field, $b_{\m\n}$, called the Kalb-Ramond field, 
which is often associated to axions in four dimensional compactifications.
\par
The coupling of these fields to the embeddings of the two-dimensional 
world sheet of the string, $\Sigma$ in the spacetime $M$, $x^{\m}(\sigma,\tau)$
is given by the two-dimensional nonlinear sigma model which in the 
conformal gauge reads:
\be
S_2 = \frac{1}{4\pi l_s^2}\int d^2 \sigma \left[ g_{\m\n}\pd_a x^{\m}\pd_b 
x^{\n}\eta^{ab}+
i b_{\m\n}\pd_a x^{\m}\pd_b x^{\n}\epsilon^{ab}+ l_s^2 R^{(2)} \phi(x)\right]
\ee
(where $\sigma^a \equiv (\sigma^1, \sigma^2)\equiv (\sigma,\tau)$ , $\eta^{ab}$
represents the two-dimensional flat metric, and $\epsilon^{ab}$ stands for the
two-dimensional Levi-Civita symbol, whereas $R^{(2)} $ stands for the
two-dimensional scalar curvature).
\par
In order for self-consistency, (i.e., two-dimensional conformal invariance,
which entails the vanishing of the corresponding $\beta$-functions) 
these backgrounds have to obey some equations which, to the lowest order 
in $l_s^2$ can be derived from a
$n$-dimensional action principle:

\be\label{string}
S_n \equiv\int d(vol)_n\frac{1}{2\kappa^2_n}~e^{- 2\phi}\left[~R 
-\frac{n-26}{3 l_s^2}-\frac{1}{12}H_{abc}H^{abc} + 4(\nabla\phi)^2\right]
\ee
(where $H\equiv d b$ is the field strength of the Kalb-Ramond field, i.e.,
$H_{abc}=\pd_{[a}b_{bc]}$, and $d(vol)_n\equiv\sqrt{g}d^n x$, whereas
$\kappa_n$ stands for the n-dimensional Planck's constant). 
\par
The mixing between the dilaton and the graviton in the kinetic term 
can be avoided with a field redefinition consistent in a Weyl transformation:
\be
g^{E}_{\m\n}\equiv e^{\frac{4(\phi_0-\phi)}{n-2}}g_{\m\n}
\ee
(where $\phi_0\equiv <\phi>$ is the unknown dilaton vacuum expectation
value,
which is different in principle from its asymptotic value at  infinity , 
$\phi_{\infty}$)
leading to the spacetime effective action written in what is usually 
called the {\em Einstein Frame} (while we say that the former action
 (\ref{string}) was written in the {\em String Frame}).
\be
S^E_{n}\equiv\int d(vol)_n\frac{1}{2\kappa^2_n}\left[~R 
-\frac{n-26}{3 l_s^2}e^{\frac{4\phi_E}{n-2}}-\frac{1}{12}
e^{-\frac{8\phi_E}{n-2}}H_{abc}H^{abc} -\frac{ 4}{n-2}(\nabla\phi_E)^2\right]
\ee
and we have defined $\phi_E\equiv \phi-\phi_0$, in such a way that $<\phi_E>=0$.
\par
There are other frames associated with topological defects, like D-branes, 
which differ in the power of the exponential of the dilaton (because they
appear at different order in the string coupling constant $g_s$). They can be
easily included in the framework of our discussion, however.
\par

Although it is not known what is the general form of the higher order 
(sigma model) corrections, it has been conjectured in \cite{damour}
that it would have the general form \footnote{This form is in any case general
  enough to include all internediate frames associated to different {\em
    branes}} in the string frame:
\bea\label{stringloop}
&&S_n=\int d(vol)_n\left[\frac{B_g(\phi)}{l_s^{n-2}} R +
\frac{B_{\phi}(\phi)}{l_s^{n-2}}(4 \nabla^2 \phi - 4(\nabla\phi)^2)\right.\nonumber\\
&&\left.-\frac{B_f (\phi)}{l_s^{n-4}}\frac{k}{4} tr F^2 - B_{\psi}(\phi)\bar{\psi}
\not\!\nabla\psi - B_m(\phi)m\bar{\psi}\psi\right]
\eea
The dilaton dressing functions $B_i (\phi)$ are unknown, and we have indicated
some of the gauge fields and fermion fields present, as well as a typical
 mass term, which will be needed in further considerations. The Kac-Moody
 level $k$ is anumerical constant.
The term involving the dilaton kinetic energy can be rewritten upon partial
 integration \footnote{Which is only valid when the dilaton field vanishes at
   infinity fast enough} as:
\be
-\frac{4}{l_s^{n-2}}(B^{\prime}_{\phi}+B_{\phi})(\nabla\phi)^2
\ee
The Einstein frame will now be defined through the Weyl transformation:
\be
g_{\m\n}\equiv \frac{(l_s m_p)^2}{B_g^{2/(n-2)}} g^E_{\m\n}
\ee
(where $m_p$ is the Plack mass) and a redefinition of the dilaton:
\be
\phi=\int^{\phi_E}
\left[\frac{n-1}{4}(\frac{B_g^{\prime}}{B_g})^2 + 
(n-2)\frac{B^{\prime}_{\phi}+B_ {\phi}}{B_g}\right]^{-1/2}
\ee
Fermions are redefined as well:
\be
\psi = \frac{B_g^{(n-1)/(2(n-2))}}{(l_s m_p)^{(n-1)/2}B_{\psi}^{1/2}}\psi_E
\ee
This leads to the Einstein frame effective action in which the Einstein
Hilbert term is not mixed with the dilaton:
\bea\label{einsteinloop}
&&S^E_n=\int d(vol)^E_n\left[\frac{1}{m_p^{2-n}} R_E 
 - \frac{4}{(n-2)m_p^{2-n}}(\nabla\phi_E)^2\right.\nonumber\\
&&\left. -\frac{B_f (\phi)}{B_g^{(n-4)/(n-2)}m_p^{4-n}}\frac{k}{4} tr F^2 - 
\bar{\psi}_E
\not\!\nabla\psi_E - B_m(\phi)m\frac{l_s m_p}{B_{\psi}B_g^{1/(n-2)}}
\bar{\psi}_E\psi_E \right]
\eea
The coefficient of the dilaton kinetic energy is conventional, and can be
fixed at will. In the new Frame there are extra terms of Yukawa type 
bewteen the dilaton and
the fermions which have been left implicit.
It is worth noticing that masses refered to the natural unit of mass in the 
string frame, i.e., $m_s\equiv\frac{B_g^{1/(n-2)}}{l_s}$ do not
change in the Einstein frame (where the unit of mass is $m_p$), provided we
divide as well by the kinetic energy factor (i.e., $B_{\psi}$). This means 
that, in a
precise sense, properly renormalized masses are invariant under change 
of frame.
\par
The two frames are usually considered as completely equivalent for describing
the physics of the massless modes of the string.
This fact (although occasionally questioned in the literature, 
(cf.\cite{capozziello},\cite{casadio},\cite{faraoni},\cite{dick})), 
seems hardly arguable, at least as long as the
(classical) functions involved are smooth; given any solution of the system
of differential equations which encode the equations of motion in
one frame, there exists a corresponding solution in the other frame.
\par
The equations of motion for the metric and the dilaton in the String Frame read:
\bea
&&\frac{\delta S}{\delta g^{\m\n}}=
\frac{B_g}{l_s^{n-2}}R_{\m\n}-\frac{4}{l_s^{n-2}}(B^{\prime}_{\phi}+B_{\phi})\nabla_{\m}\phi\nabla_{\n}\phi
-\frac{B_f}{l_s^{n-4}}\frac{k}{2}tr
F_{\m\a}F_{\n}^{\a}-B_{\psi}\bar{\psi}\gamma_{\m}\nabla_{\n}\psi \nonumber\\
&&-\frac{1}{2}g_{\m\n}(\frac{B_g}{l_s^{n-2}} R -
\frac{4(B_{\phi}+B_{\phi}^{\prime})}{l_s^{n-2}}(\nabla\phi)^2
-\frac{B_f }{l_s^{n-4}}\frac{k}{4} tr F^2 - 
B_{\psi}\bar{\psi}\not\!\nabla\psi -m  B_m 
\bar{\psi}\psi )=0
\eea
and
\bea
&&\frac{\delta S}{\delta\phi}=\frac{B^{\prime}_{\phi}}{l_s^{n-2}} R
  +\frac{4}{l_s^2}(B_{\phi}^{\prime\prime}+B_{\phi}^{\prime})(\nabla\phi)^2-
\frac{k B_f^{\prime}}{4 l_s^{n-4}}tr F^2\nonumber\\
&& - B_{\psi}^{\prime}\bar{\psi}\not\!\nabla\psi
      +\frac{8}{l_s^{n-2}}(B^{\prime}_{\phi}+B_{\phi})\nabla^2\phi
      -m B_m^{\prime}\bar{\psi}\psi=0.
\eea
The equivalence of the equations of motion in the two frames has been worked
out in detail in (\cite{campbell}).
Let us now however discuss in turn some delicate issues.

\section{Dualities}

There is now a certain amount of evidence for different
kinds of symmetries between different string theories (See for example
\cite{hull}). The two more important ones are {\em S-duality} and
 {\em T-duality}. We shall say that two (not neccessarily different)
theories, $T_1$ and $T_2$ are T-dual, when $T_1$ compactified at large
Kaluza-Klein volume is physically equivalent to $T_2$ at small
Kaluza-Klein volume. If we call $t$ the modulus associated to global variations
of the Kaluza-Klein volume, by $Vol \sim e^t$, this implies a 
relationship of the general form
\be
t(1) \; =\;  - t(2) \; .
\ee
This symmetry can be proven true when there is an isometry in the spacetime
manifold whose Killing vector is written in adapted coordinates as
$\frac{\pd}{\pd x^0}$
by several means
(\cite{alvarez},\cite{giveon} )and is such that the string metric transforms in
a simple way, to wit:
\bea
{\tilde g}_{00}&=&{1\over g_{00}} \; ,\nonumber\\
         {\tilde g}_{0i}&=&{b_{0i} \over g_{00}} \; ,
\qquad
{\tilde b}_{0i}={g_{0i} \over g_{00}} \; ,\nonumber\\
          {\tilde g}_{ij} &=& g_{ij} -
{g_{0i}g_{0j} - b_{0i} b_{0j}\over g_{00}} \; ,
\nonumber\\
        {\tilde
b}_{ij}&=&b_{ij}-{g_{0i}b_{0j}
         -g_{0j}b_{0i}\over g_{00}} \; .
\eea

\par
S-duality, on the other hand, refers to 
the equivalence of $T_1$ at small coupling
with $T_2$ at large coupling. Given the relationship we already mentioned
between the dilaton and the string coupling, it demands for the 
dilaton something like
\be
\phi (1) \; = \; - \phi (2) \; ,
\ee
and, by definition, lies beyond the possibilities of verification by 
means of perturbation theory. 
For $IIB$ strings, the (modified, cf. ref. (\cite{tomas}) Einstein metric is inert under this transformation,
\be
\tilde{g}^{E}_{\m\n}=g^{E}_{\m\n}
\ee
\par

We then se that each frame seems to be  most appropiate depending on 
which symmetry we
believe to be the most fundamental.


\section{The definition of the vacuum state}
Insofar as
\be
T_{\m\n}^{(matt)}=0
\ee
we always have a {\em vacuum} solution of the equations og motion:
\bea
&&\phi=0\nonumber\\
&&g_{\m\n}=\eta_{\m\n}
\eea
On the other hand, it has been proven in ref. \cite{damour} that under certain
hypothesis (mainly the equality of all dressing functions $B_i$), the
cosmological evolution attracts the dilaton towards the point where the
dressed masses are stationary
\be
\frac{\pd m_E}{\pd \phi}|_{\phi=\phi_0}=0
\ee
It has also been argued that the existence of several different minima is
probably incompatible with present bounds on the equivalence principle (namely
a relative difference in acceleration $\frac{\Delta a}{a}\leq 10^{-13}$
 \cite{damour})).
Although it is known that in general the hypothesis of reference
(\cite{damour}) are not fulfilled, there are actual string models where 
this mechanism is automatic 
(\cite{antoniadis}).
\section{The principle of equivalence}
Were to be strings the probes of the metric, it is obvious that the most
natural frame would be the String Frame. But in most classical experiments,
the metric is detected through its effect on classical test particles, which 
describe geodesics of the spacetime metric. At a higher level of precision,
the geodesic deviation equation, gives direct information of the 
Riemann tensor.
\par
Let us now review how the concept of {\em particle} is recovered from the
concept of {\em field}. The latter is written as a formal (WKB) series
\be 
 \phi = e^{\frac{1}{\epsilon}\sum \epsilon^n \phi_n}
\ee
The Klein-Gordon equation then gives, in the eikonal approximation (actually,
 to the dominant order $\frac{1}{\epsilon^2}$)
\be
k^2 = - m^2
\ee
where the mass has to be considered as $o(\epsilon^{-2})$, and
\be
k_{\m}\equiv\nabla_{\m}\phi_0
\ee
This implies that the flow lines defined by the congruence $k^{\m}$ are
geodesic, since
\be
0=\nabla_{\rho}(k^2)=2 k^{\m}\nabla_{\rho}k_{\m}=  2 k^{\m}\nabla_{\m}k_{\rho}
\ee
where the last step is justified since the vector $k^{\rho}$ is itself a 
gradient.
\par
Let us now consider a scalar field other than the dilaton (which remains
massless to all orders in perturbation theory), with Lagrange density:
\be
L\equiv -\frac{1}{2}(B_{\chi} (\nabla\chi)^2+ m^2 B_m \chi^2)
\ee
The equations of motion in the eikonal approximation now yield
\be
B_{\chi}k^2 = - m^2 B_m
\ee
which violates the principle of equivalence unless $B_m=B_{\chi}$, but does
    not distinguish qualitatively (although it does it quantitatively) between
    different frames. 
\par
That is, particles will propagate along geodesics of
    that metric (if any) such that $B_{\chi}=B_m$. In addition, the 
dressing factors $B_{\chi}$ will depend generically on the particle
    considered,
and so will depend the trajectories, and it is this fact which violates the
    equivalence principle.
\section{The fluid approximation}
The situation is perhaps less clear when both the metric and the dilaton
are singular in one frame, but the metric is regular in the other.
This clearly changes the physics of {\em test particles} propagating in 
the physical spacetime. 
\par
Let us now point out a related, but simpler, situation.
\par
Imagine that matter (that is all fields except gravitation itself)is such
that its energy-momentum tensor corresponds to a perfect fluid,
\be
T_{\m\n}\equiv (\rho + p)~u_{\m}u_{\n} + p g_{\m\n}
\ee
Then the question is: will the same matter still behave as a perfect 
fluid in the other frame? This is a meaningful question, because matter
is almost always considered of such a form in cosmological investigations.

The general conditions for a fluid description to be valid 
consist in demanding that the wavelength  (as measured in a 
Local Inertial Frame, LIF, defined by a {\em vielbein}, 
$e_a^{\m}\pd_{\m}~,a=0,\ldots,n-1$) should be much less than both the 
macroscopical length of the wavepacket, $l$, and the scale of variation of 
Riemann's tensor, $r$.
\be
\lambda<< ~min ~(r,l)
\ee
Now, when changing frames, quantities in the LIF obviously do not
change,(neglecting the new interacions introduced through the change of frame)
 whereas lengths scale as
\be\label{l}
l_S = l_E \frac{l_s m_p}{B_g^{1/(n-2)}}
\ee
(Where an average value for the dilaton field in the region considered is
implicitly  assumed).
In the case of Riemann's tensor, there are in addition extra terms
proportional to the square of the first derivative and to the second derivative
of the dilaton field, which could dominate for a rapidly fluctuating dilaton.

A sufficient condition for such a hydrodynamic fluid description is to be at
 thermodynamic equilibrium. It is in turn clear that a neccessary
 condition
for it  (it is less clear whether it will be {\em sufficient}) 
is that the mean free time between collisions 
should be smaller than Hubble's time
\be
\tau << H^{-1}
\ee
The mean free time, in turn, is related to quantities computed in a LIF:
basically $\tau\sim (\rho v \sigma)^{-1}$, where $\rho$ is the average
density of particles, $v$ is the average velocity, and $\sigma$ is the
total cross section (again, neglecting the new interactions).
\par
The scale factor, on the other hand, is a global quantity, and, as such,
scales as above
\be
R_S = R_E \frac{l_s m_p}{B_g^{1/(n-2)}}
\ee
in such a way that
\be
H_S\equiv \frac{\dot{R}_S}{R_S} = H_E - \frac{1}{n-2}\frac{B_g^{\prime}}{B_g}
\dot{\phi}
\ee
The crutial quantity now is the dimensionless
quotient
\be
\frac{B_g^{\prime}\dot{\Phi}}{(n-2) B_g H_E}
\ee
If it is small, equilibrium in both frames is equivalent.
\par 
But if it is large (corresponding physically to a wildly fluctuating dilaton
{\em or} to the vicinity of a place in which the dilaton itself is singular),
then equilibrium in Einstein's frame does not guarantee equilibrium 
in the string frame.
\par
It is worth remarking that even for a time independent dilaton (which
does not spoil the equilibrium condition) the preceding formula (\ref{l})
indicates that for large dilatons averaged (corresponding to large $g_s$),
so that $B_g<<1$ and correspondingly, $l_s>>l_E$, 
 matter enjoying a 
fluid description in the String Frame does not neccessarily do so 
in the Einstein Frame. For large {\em negative} averaged dilaton couplings
(corresponding to small $g_s$) the converse is true: a fluid description in
the Einstein Frame does not guarantee a fluid description in String Frame).
\section{Conclusions}
There is in our opinion no doubt of the equivalence of all frames for the
description of the gravitational effects of string theories at a basic level,
at least when all functions involved are smooth.
\par
When this is not the case, the solution depends on what is the quantum
resolution of the classical gravitational singularities. The symmetries of
string theory (T-duality, in particular) suggest that there is, in a sense, a
minimal measurable length, but the issue is far from settled; it could be, in
particular, that certain particle-like topological defects, known as $D0$
branes, could probe shorter lengths.
\par
A different question is what is the classical metric felt by a particular 
probe (usually, a {\em test particle}). Here, again, strings give a unique
answer (depending on the probe used), which seems difficult to
reconcile {\em a priori} with existing bounds on violations of the 
equivalence principle. A
detailed comparison is however difficult owing to our limited understanding of
the dynamics of the dilaton as well as other scalars in the string spectrum.

\section*{Acknowledgments}
We are indebted to Alberto Casas, Pilar Hern\'andez, Tom\'as Ort\'{\i}n and 
Gabriele Veneziano for illuminating discussions.
This work ~~has been partially supported by the
European Commission (HPRN-CT-200-00148) and CICYT (Spain).


\appendix


\end{document}